# MUFold-BetaTurn: A Deep Dense Inception Network for Protein Beta-Turn Prediction

Chao Fang[1], Yi Shang[1, *] and Dong Xu[1, 2, *]

[1]Department of Electrical Engineering and Computer Science, University of Missouri, Columbia, Missouri 65211, USA, [2]Christopher S. Bond Life Sciences Center, University of Missouri, Columbia, Missouri 65211, USA.

*To whom correspondence should be addressed: shangy@missouri.edu, xudong@missouri.edu

## Abstract
Beta-turn prediction is useful in protein function studies and experimental design. Although recent approaches using machine-learning techniques such as SVM, neural networks, and K-NN have achieved good results for beta-turn prediction, there is still significant room for improvement. As previous predictors utilized features in a sliding window of 4–20 residues to capture interactions among sequentially neighboring residues, such feature engineering may result in incomplete or biased features, and neglect interactions among long-range residues. Deep neural networks provide a new opportunity to address these issues. Here, we proposed a deep dense inception network (DeepDIN) for beta-turn prediction, which takes advantages of the state-of-the-art deep neural network design of the DenseNet and the inception network. A test on a recent BT6376 benchmark shows that the DeepDIN outperformed the previous best BetaTPred3 significantly in both the overall prediction accuracy and the nine-type beta-turn classification. A tool, called MUFold-BetaTurn, was developed, which is the first beta-turn prediction tool utilizing deep neural networks. The tool can be downloaded at http://dslsrv8.cs.missouri.edu/~cf797/MUFoldBetaTurn/download.html.

## 1 Introduction

Protein tertiary structure prediction is an important and challenging problem, which has been an active research topic for the past 50 years (Dill and MacCallum, 2012; Zhou et al., 2011; Webb and Sali, 2014). Since it is challenging to predict the protein tertiary structure directly from a sequence, this problem has been divided into small sub-problems, such as protein secondary structure prediction. The protein secondary structure can be divided into three classes: alpha-helix, beta-sheets and coil (Richardson, 1981). The coil region can be classified as tight turns, bulges and random coil structures (Milner-White et al., 1987). The tight turns can be further classified into alpha-turns, beta-turns, gamma-turns, delta-turns, and pi-turns (Rose et al., 1985). Among these tights turns, beta-turns represent the most abundant type in proteins. For example, in the BT6376 (Singh et al., 2015) data set, we found 126,016 beta-turns (9%) out of 1,397,857 amino acids. By definition, a beta-turn contains four consecutive residues (denoted by $i$, $i+1$, $i+2$, $i+3$) if the distance between the $C\alpha$ atom of residue $i$ and the $C\alpha$ atom of residue $i+3$ is less than 7 Å and if the central two residues are not helical (Lewis et al., 1973). An alternative but more precise definition of beta-turn is the possession of an intra-mainchain hydrogen bond between the CO of residue i and the NH of residue i+3 (Chou 2000) (see Figure 1 for an illustration of what a beta-turn is). There are nine types of beta-turns, which are classified based on the dihedral angles of two central residues in a turn (Hutchinson and Thornton, 1994) as shown in Table 1. Beta-turns can be assigned from a PDB structure by using the PROMOTIF software (Hutchinson and Thornton, 1994). Beta-turns play an important role in mediating interactions between peptide ligands and their receptors (Li et al., 1999). In protein design, loop segments and hairpins can be formed by introducing beta-turns in proteins and peptides (Ramirez-Alvarado et al., 1997). Hence, it is important to predict beta-turns from a protein sequence (Kaur et al., 2002).

**Table 1.** Nine types of beta-turns and their dihedral angles of central residues in degrees

| Turn Type | $Phi_1$ | $Psi_1$ | $Phi_2$ | $Psi_2$ |
|---|---|---|---|---|
| I | -60 | -30 | -90 | 0 |
| I | 60 | 30 | 90 | 0 |
| II | -60 | 120 | 80 | 0 |
| II' | 60 | -120 | -80 | 0 |
| IV | -61 | 10 | -53 | 17 |
| VIII | -60 | -30 | -120 | 120 |
| VIb | -135 | 135 | -75 | 160 |
| VIa1 | -60 | 120 | -90 | 0 |
| VIa2 | -120 | 120 | -60 | 0 |

The locations of $Phi_1$, $Psi_1$, $Phi_2$ and $Psi_2$ are illustrated in Figure 1.

The early predictors (Hutchinson and Thornton, 1994; Chou and Fasman 1974; Chou and Fasman, 1979) used statistical information derived from protein tertiary structures to predict beta-turns. The statistical methods were based on the positional frequencies of amino acid residues. Zhang and Chou (1997) further observed the pairing of the first and fourth residues, and the second and the third residues, plays an important role in beta-turn formation. They proposed the 1-4 and 2-3 correlation model to predict beta-turns (Zhang and Chou, 1997). Later, Chou (1997) applied a sequence-coupled approach based on the first-order Markov chain to further improve their prediction model. Kaur and Raghava (2002) developed a web server, called BetaTPred, which implemented this model and achieved a Matthew Correlation Coefficient (MCC) of 0.26 in beta-turn prediction.

McGregor *et al.* (1989) used neural networks to predict beta-turns, which is the first machine-learning approach for beta-turn prediction, and they achieved an MCC of 0.20. Shepherd *et al.* (1999) developed BTPred using secondary structure information as input and achieved an MCC of 0.34. Kim (2004) applied a K-nearest neighbor method for beta-turn prediction and improved MCC to 0.40. Fuchs and Alix (2005) further improved the MCC to 0.42 by incorporating multiple features such as propensities, secondary structures and position specific scoring matrix (PSSM). Kaur and Raghava (2004) developed the BetaTPred2 server, which used a two-layer neural network with an MCC of up to 0.43. Kirschner and Frishman (2008) developed MOLEBRNN using a novel bi-directional recurrent neural network (Bi-RNN), with an MCC of up to 0.45. Hu and Li (2008) used a support vector machine (SVM) and incorporated features such as increment of diversity, position conservation scoring function and secondary structure to raise the MCC up to 0.47. Zheng and Kurgan (2008) used the predicted secondary structure from PSIPRED (Jones 1999), JNET (Cole *et al.*, 2008), TRANSEEC (Montgomerie *et al.*, 2006), and PROTEUS2 (Montgomerie *et al.*, 2008) to improve the performance. Kountouris and Hirst (2010) used predicted dihedral angles along with PSSM and predicted secondary structures to achieve MCC of 0.49. Petersen *et al.* (2010) developed the NetTurnP server with an MCC of 0.50 by using independent four models for predicting four positions in a beta-turn. Singh *et al.* (2015) developed the Be-

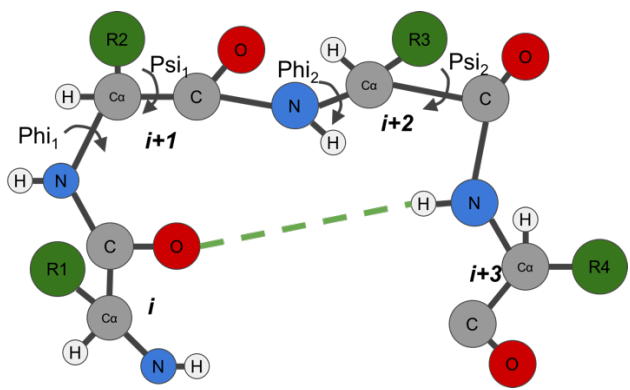

**Fig. 1.** An illustration of what a beta-turn is. C, O, N, and H, represent carbon, oxygen, nitrogen and hydrogen atoms, respectively. R represents side chain. The dashed line represents the hydrogen bond.

taTPred3 server to achieve MCC of 0.51 using a random forest method, which was the most accurate method before our work.

The above-mentioned machine-learning methods achieved some successful results in beta-turn prediction. However, there is significant room for improvement, particularly in predicting nine types of beta-turns. Most of these previous methods relied on a sliding window of 4-10 amino acid residues to capture short interactions. Also, previous neural networks with one or two layers (shallow neural networks) could not extract high-level features from input data sets. So far, no deep neural networks have been applied to beta-turn prediction. Deep neural networks can learn representations of data with multiple levels of abstraction (LeCun et al., 2015), which provides a new opportunity to this old research problem.

Here, a deep neural network, named DeepDIN, was proposed for beta-turn prediction. The contributions are presented as follows: 1) MUFold-BetaTurn is the first beta-turn prediction software to utilize the deep-learning framework and outperformed the previous best predictor BetaTPred3 (Singh *et al.*, 2015); 2) we employed strategies such as balanced learning and transfer learning to tackle the problem of small sizes and imbalanced data sets for deep learning, which may provide a good example for some other deep-learning applications in bioinformatics; and 3) we provide a free standalone software for the research community to use.

## 2 Methods and Materials

### 2.1 Preliminaries and Problem Formulation

To make an accurate prediction, it is important to provide useful input features to machine-learning models. In our method, we carefully designed feature matrices corresponding to the primary amino acid sequence of a protein. Specifically, feature sets include:1) a physicochemical feature set describing properties of amino acids, 2) an HHBlits profile, 3) prediction of eight state secondary structures from MUFold-SS (Fang *et al.*, 2018a) and 4) a shape string predicted by Frag1D (Zhou *et al.*, 2009).

Physicochemical features describe hydrophobic, steric, and electric properties of amino acids and provide useful information for protein sequence analysis and prediction. By using physicochemical properties, protein sequences are represented as an informative dense matrix. The physicochemical feature matrix consists of seven physico-chemical properties as defined by Heffernan *et al.* (2017), plus a number 0 or 1 representing the existence of an amino acid at this position as an input (called *NoSeq* label). The reason for adding the *NoSeq* label is because the proposed deep neural networks are designed to take a fixed-size input, such as a sequence length of 700 residues in our experiment. To run a protein sequence shorter than 700 through the network, the protein sequence will be padded at the end with 0 values and the *NoSeq* label is set to 1. If the protein is longer than 700 residues, it can be split into multiple segments, each shorter than 700 residues. Hence, a protein sequence will be represented as a 700-by-8 matrix, which is the first input feature for DeepDIN.

The second set of useful features comes from the protein profiles generated using HHBlits (Remmert *et al.*, 2012). In our experiments, the HHBlits software used the database uniprot20_2013_03, which can be downloaded from http://wwwuser.gwdg.de/~compbiol/data/hhsuite/databases/hhsuite_dbs/. The profile values were transformed by the sigmoid function into the range (0, 1). Each amino acid in the protein sequence is represented as a vector of 31 real numbers, of which 30 are from amino acids HHBlits profile values and one is a *NoSeq* label in the last column. The HHBlits profile contains amino acids and some transition probabilities: "A, C, D, E, F, G, H, I, K, L, M, N, P, Q, R, S, T, V, W, Y, M->M, M->I, M->D, I->M, I->I, D->M, D->D, Neff, Neff_I, and Neff_D". HHBlits' profiles are more sensitive than PSI-BLAST profiles and provide useful evolutionary features for the protein sequence. A HHBlits profile is represented as a 700-by-30 matrix, which is the second input feature for DeepDIN.

The third set of useful features, the predicted shape string, comes from Frag1D (Zhou *et al.*, 2009). For each protein sequence, the Frag1D can predict useful protein 1D structure features: the classical three-state secondary structure, three- and eight-state shape string. A classical three-state secondary structure contains an H (helix), S (sheet), and R (random loop). Eight state shape string labels are defined: R (polyproline type alpha structure), S (beta sheet), U, V (bridging regions), A (alpha helices), K ($3_{10}$ helices), G (almost entirely glycine), and T (turns). Shape strings (Ison *et al.* 2005) can describe a 1D string of symbols representing the protein backbone Psi-Phi torsion angles. They include regular secondary structure elements; thus, shape 'A' corresponds to alpha helix and shape 'S' corresponds to beta sheets. In addition, shape strings classify the random loop regions into several states that contain much more conformation information. For the Frag1D predicted result, each amino acid in the protein sequence is represented as a vector of 15 numbers: Three are from the classical three-state secondary structures, three are from the three-state shape strings, eight are from the eight-state shape strings, and one *NoSeq* label is in the last column. The predicted classical three-state secondary structure feature is represented as one-hot encoding as follows: helix: (1,0,0), strand: (0,1,0), and loop: (0,0,1). The same rule applies to three- and eight-state shape string features. Hence, a Frag1D result will be represented as a 700-by-15 matrix, which is the third input feature for DeepDIN.

The fourth set of useful features comes from our previous designed secondary structure prediction tool: MUFold-SS (Fang *et al.*, 2018a). MUFold-SS achieved state-of-the-art performance in an eight-state secondary structure prediction and it should provide useful features in beta-turn prediction tasks. The eight-state secondary structure has the following components: H (alpha helix), B (beta bridge), E (extended strand), G (3-helix), I (5 helix), T (hydrogen bonded turn), and S (bend). Hence, an eight-state predicted secondary structure will be represented as a 700-by-8 matrix, which is the fourth input feature for DeepDIN.

Protein beta-turn prediction is a classification problem. To be specific, it can be formulated as a residue-level prediction or turn-level prediction as first proposed by (Singh *et al.*, 2015).

- **Residue-level prediction**: To predict whether each amino acid has a class label of a turn or non-turn. Here, the predicted output is a 700-by-3 matrix, where '700' is the sequence length 700 and '3' stands for two-state labels plus 1 *NoSeq* label.
- **Turn-level prediction**: At the turn-level, a sliding window of four residues was used to generate the turn-level data sets. And the overall predicted beta-turn output of a protein sequence is represented as a fixed-size matrix, with a (700-4+1) dimension by 3 for two-state labels plus 1 *NoSeq* label matrix. For a nine-class classification problem, the label is: turn of a specific type or non-turn.

The evaluation metric for beta-turn prediction typically uses MCC more often than accuracy, since the accuracy only considers the true positive and false positive without the true negative and false negative, and non-beta-turns (negative data) dominate the data. MCC can be calculated as follows:

$$MCC = \frac{TP*TN - FP*FN}{\sqrt{(TP+FP)(TP+FN)(TN+FP)(TN+FN)}} \quad (1)$$

where TP is the number of true positives, TN is the number of true negatives, FP is the number of false positives and FN is the number of false negatives.

### 2.2 New Deep Dense Inception Networks for Protein Beta-Turn Prediction (DeepDIN)

In this section, a new deep dense inception network architecture (DeepDIN) is presented. The architecture makes use of deep inception (Szegedy *et al.*, 2017) networks and dense network (Huang *et al.*, 2017). Figure 2 (A) presents our proposed network design. The overall beta-turn prediction pipeline contains the following steps:

1. Given a protein sequence, generate four sets of features: physicochemical feature, profile features from HHBlits, predicted shape string (using Frag1D) and predicted eight-state secondary structure (using MUFold-SS).
2. Perform the convolution operation on each feature to get the convolved feature map.
3. Concatenate four convolved feature maps along the feature dimension.

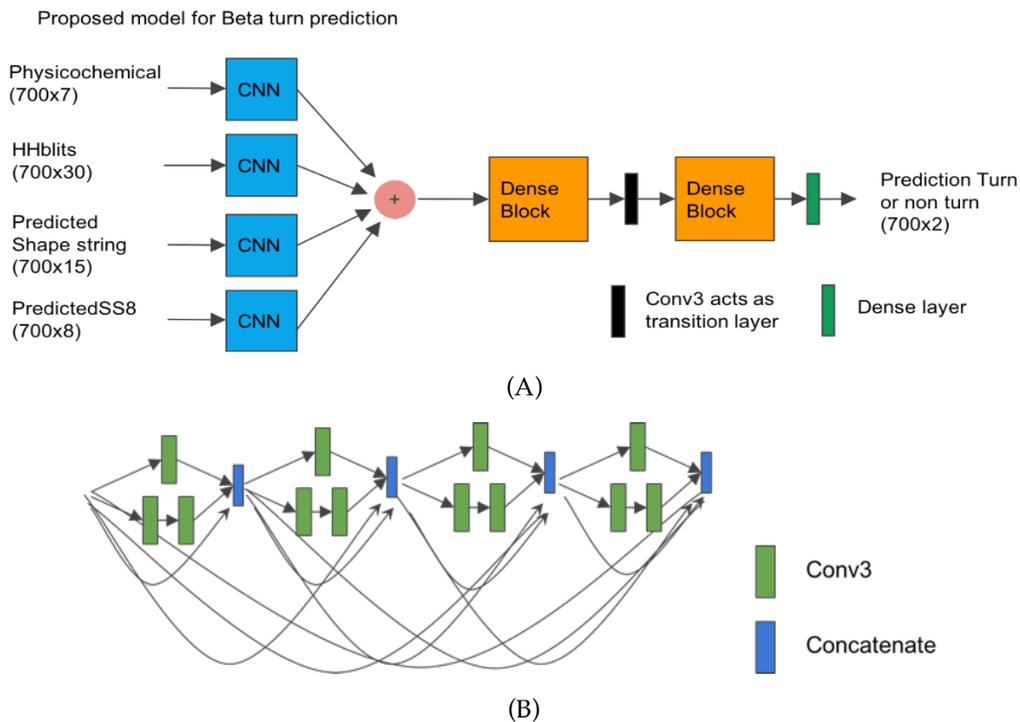

**Fig. 2.** Network architecture. (A) shows the overall proposed model for beta-turn prediction. (B) shows the details of a dense inception module that consists of four inception modules, each of which is fully connected in a feed forward fashion.

4. Feed the concatenated feature map into the stringed dense inception blocks. In between there is a convolutional layer acting as the transition layer.
5. Predict beta-turn (either turn or non-turn) in the last dense layer, which uses Softmax as the activation function.

Figure 2 (B) shows details of a dense inception block. It consists of four small version of inception blocks and each inception block is fully connected with other inception blocks. In other words, a dense inception module is constituted by connecting each inception layer to every other inception layer in a feed-forward fashion. The design of dense inception modules can extract non-local interactions of residues over a diverse range in a more effective way. Adding more dense inception blocks is possible but requires more memory.

Each convolution layer, such as 'Conv (3)' in Figure 2, consists of four operations sequentially: 1) a one-dimensional convolution operation using the kernel size of three; 2) the batch normalization technique (Ioffe and Szegedy, 2015) for speeding up the training process and acting as a regularizer; 3) the activation operation, ReLU (Radford et al., 2015); and 4) the dropout operation (Srivastava et al., 2014) to prevent the neural network from overfitting by randomly dropping neurons during the deep network training process so that the network can avoid or reduce co-adapting. DeepDIN was implemented, trained, and tested using TensorFlow and Keras. All the experiments were performed on an Alienware Area-51 desktop computer equipped with Nvidia Titan-X GPU (11 GB memory). In our experiments, many network parameters and training parameters were tried. For the results reported, the dropout rate was set at 0.4. The optimizer used during training is Adam (Kingma and Ba, 2014), which can control the learning rate dynamically for network weight updates. There are nine beta-turn classifications, and the observations for a certain class can be as many as 40,000 or as little as 100. In different classification tasks, the batch size varies from 50 to 200. The maximum number of epochs is set up to 100. The training time varies from 2 hours to 5 hours depending on the data size when training different classification models.

Our previous studies (Fang et al., 2017; Fang et al, 2018a; Fang et al., 2018b) have successfully applied the stacked convolutional neural network (CNN), the inception module (Szegedy et al., 2017), the residual module (He et al., 2016) to protein sequence analysis and prediction problems. Following our previous work, here we propose a new deep neural network architecture called dense inception network (DeepDIN) for beta-turn prediction.

## 2.3 Apply Transfer Learning and Balanced Learning to DeepDIN to Handle Imbalanced Small Data set

Deep learning usually requires a large amount of data sets to train the network. Here, since some of the beta-turn class has only a small amount of data set, the following deep-learning strategy was used to address the small data sets classification problem.

**Balanced learning**: Many previous researchers have studied the problem of imbalanced data learning and proposed ways of balancing the data set by using either over-sampling like SMOTE (Han et al., 2015) or under-sampling like Tomek Links (Tomek, 1976). The beta-turn data is highly unbalanced, and this would cause a deep neural network like DeepDIN prone to predict everything as in the negative class. In this experiment, the protein beta-turn data classified as positive data are not very large; thus, any up-sampling or down-sampling may cause loss or pollution of useful information. Rather than resampling the training data points, a balanced learning approach was adapted to handle the problem. The weighted cross entropy was used as the loss function, which is defined as the following functions:

$$\forall i \in [0,N), H(i) = -\sum_i y_{i,l} * \log \hat{y}_{i,l} \quad (2)$$

$$H = \sum_l W * y_i * H(i) \quad (3)$$

where $l$ is the number of classes; $W = (w_1, w_2, ..., w_l)$ is the weight for each class; $N$ is the total number of data points; $y_i$ is the one-hot ground truth label; $\hat{y}_i$ is the predicted probability distribution of a data point; $H(i)$ is the cross-entropy loss of one data; and $H$ is the weighted cross-entropy. The weighted cross entropy as the loss function was able to address the issue caused by the small sample size by rescaling the prediction of each class by its weight. In this implementation, the class weights were calculated using the training data and assigned using the Scikit-learn (Pedregosa et al., 2011) toolbox. The balanced class weights are given by $n\_samples/(n\_classes * bincount(y))$, where $y$ is an array of original class labels per sample, and "bincount" is a built-in function from the Scikit-learn toolbox to calculate the number of bins.

**Transfer Learning**: We applied transfer learning to handle the limited number of beta-turn data used in the training set. The idea of transfer learning originally came from the image classification problem (Raina et al., 2007). This technique was proposed to handle the insufficient size of a data set that can be used to train the deep neural network. In our study, since there are nine classes of beta-turns, and especially since those in VIa1, VIa2, and VIb contain a few hundred data points, the amount of data belonging to each class may not produce a model with the ability to extract features or to be well generalized. To solve this problem, the pre-trained weights from DeepDIN are used to classify two-class beta-turns, which were each loaded into a separate nine-class classification model as initial weights. Notably, the pre-trained model here is the beta-turn model used to classify the two-class beta-turn problem. Since that model has "observed" some generic features of what a beta-turn is, it should be useful to many specific nine-class beta-turn classification tasks. Then, to train each individual class model, each individual class training set was used to further train the pre-trained network. The weights of the pre-trained networks were fine-tuned by continuing the backpropagation. It is possible to fix the earlier layers of the deep networks due to the overfitting issue and only fine-tune the high-level portion of the network. After the trials, the overall network was fine-tuned without freezing the weights from lower level of the network.

For the learning rates during the transfer learning, a smaller learning rate (0.005) and batch size (10) were used to train the network and fine-tune the network weights in the fine-tuning. The reason is that the pre-trained network weights are relatively good, a slower and smaller learning rate will not distort them too quickly and too much before it converges.

## 2.4 Benchmark data sets

The following two publicly available benchmark data sets were used in our experiments:

1) **BT426** (Guruprasad and Rajkumar, 2000) is a data set commonly used for benchmarking beta-turn prediction methods. BT426 contains 426 protein chains in total with 25% sequence identity cutoffs, and X-ray structures of a resolution better than 2.0 Å. This benchmark was used to compare the performance of many previous predictors, such as BetaTPred3 (Singh et al., 2015) and NetTurnP (Pe-tersen *et al.*, 2010). In this work, five-fold cross-validation experiments on BT426 were performed and results were compared against other predictors.

2) **BT6376** (Singh *et al.*, 2015) is a public benchmark containing 6376 non-homologous protein chains. No two protein chains have more than 30% sequence identity. The structures of these proteins were determined by X-ray crystallography at 2.0 Å resolution or better. Each chain contains at least one beta-turn. The beta-turn labels were assigned by the PROMOTIF program (Hutchinson and Thornton, 1996). This benchmark provides data sets for both the two-class beta-turn classification and the nine-class beta-turn classification. For the nine-class beta-turn classification, labels were annotated by using PROMOTIF. Table 2 shows a list of class sizes on the samples, which indicates that the data set is very imbalanced, as non-turns occupy most of the data samples.

**Table 2.** Nine-class beta-turn and non-turn class sizes in BT6376

| Beta-turn types | # of proteins | # of Turns | # of non-turns | turns / non-turns |
|---|---|---|---|---|
| Type I | 6039 | 42,393 | 1,366,911 | 0.031 |
| Type I' | 2786 | 4353 | 747,596 | 0.005 |
| Type II | 4750 | 13,559 | 1,183,457 | 0.011 |
| Type II' | 1995 | 2643 | 545,300 | 0.004 |
| Type IV | 5950 | 38,201 | 1,360,907 | 0.028 |
| Type VIa1 | 600 | 654 | 182,122 | 0.003 |
| Type VIa2 | 177 | 188 | 56,761 | 0.003 |
| Type VIb | 914 | 1082 | 263,099 | 0.004 |
| Type VIII | 4257 | 10,111 | 1,114,707 | 0.009 |

## 3 Results and Discussion

In this section, extensive experimental results of the proposed deep neural networks on the benchmark data sets and performance comparison with existing methods are presented. To evaluate the performance of our tool, a five-fold cross validation technique was used on all data sets.

**Table 3.** Effects of feature combinations on prediction performance

| physico-chemical | SS8 | Shape string | HHBlits profile | MCC |
|---|---|---|---|---|
| x | | | | 0.186 (±0.003) |
| | x | | | 0.321 (±0.004) |
| | | x | | 0.372 (±0.004) |
| | | | X | 0.326 (±0.002) |
| x | x | | | 0.406 (±0.003) |
| x | | x | | 0.400 (±0.002) |
| x | | | X | 0.366 (±0.001) |
| | x | | X | 0.369 (±0.002) |
| | x | x | | 0.435 (±0.006) |
| | | x | X | 0.421 (±0.003) |
| | x | x | X | 0.451 (±0.002) |
| x | x | x | | 0.465 (±0.002) |
| x | | x | X | 0.440 (±0.002) |
| x | x | | X | 0.416 (±0.002) |
| x | x | x | X | 0.469 (±0.002) |

### 3.1 How Features Affect the DeepDIN Performance

Since we used four features for our proposed DeepDIN architecture, it is important to quantitatively determine how much improvement the proposed DeepDIN can make by using single, some, or all of the features (see Table 3). We used Type I beta-turns for experiments. This data set has 6,039 proteins containing a total of 42,393 Type I beta-turns. In each experiment, five rounds of five-fold cross-validation was performed to take into consideration data variation and model variation. The running time for each experiment was around 15 to 17 hours.

Table 3 shows that the predicted shape string feature either used alone or used in combined with other features can highly improve the prediction performance. Notably, some combined features have better prediction results than those features used alone, which means features in-between can have some complementary affect. For instance, when the predicted shape string is combined with the predicted secondary structure as input, the prediction performance is much better than just using the predicted shape string. Although both features are related to the protein secondary structure, they may capture different aspects of features for a protein.

### 3.2 Residue-Level and Turn-Level Two-Class Prediction on BT426

**Table 4.** Residue-Level Prediction on BT426

| Predictor | MCC |
|---|---|
| BTPRED (Shepherd *et al.*, 1999) | 0.35 |
| BetaTPred2 (Kaur and Raghava, 2002, 2004) | 0.43 |
| Hu and Li (2008) | 0.47 |
| NetTurnP (Petersen *et al.*, 2010) | 0.50 |
| BetaTPred3-Tweak (Singh *et al.*, 2015) | 0.50 |
| BetaTPred3-7Fold (Singh *et al.*, 2015) | 0.51 |
| BetaTPred3 (Singh *et al.*, 2015) | 0.51 |
| DeepDIN | **0.647 (±0.016)** |

Table 4 shows experiment results of comparing DeepDIN with existing methods at residue-level on benchmark BT426. Other than BetaTPred3 (Singh *et al.*, 2015), all other tools only had residue-level predictions. At turn-level prediction, BetaTPred3 (Singh *et al.*, 2015) achieved MCC 0.43; our tool DeepDIN achieved 0.550 (±0.019). At both residue-level and turn-level, DeepDIN outperformed all existing methods. It is noted that in (Tang *et al.*, 2011), the authors reported their turn-level MCC around 0.66 on benchmark BT426. However, they preprocessed the data set in a way that the negative (non-turn) samples were randomly selected when the ratio of positive to negative was 1:3 (Tang *et al.*, 2011), which significantly favored MCC over the setting used by BetaTPred3 and DeepDIN, i.e. using all the residues in the original BT426 data set with the ratio of positive to negative as high as 1:9. In addition, the study by Tang *et al.*, (2011) did not have a downloadable software tool or web-server to compare with. Hence, we did not compare the performance of Tang *et al.* (2011) in this study.

### 3.3 Turn-level Nine-Class Prediction on BT6376

Table 5 shows the nine-class beta-turn prediction results. The average MCC results of DeepDIN for large class beta-turns such as I, I', II, II', IV, outperformed BetaTPred3 by at least 7%, which is a significant improvement. For some small classes such as VIa1, VIb the improvement is about 3%. DeepDIN performs not as good as BetaTPred3 in other small classes such as VIa2 in average MCC, particularly because of the small amount of training data points available. Generally speaking, deep neural networks are biased to predicted negative classes if a very small amount of imbalanced observations are given. However, by applying balanced learning and transfer learning, this issue can be alleviated or overcome.

Since the beta-turn data set is very imbalanced, during training, different class weights were calculated using the Scikit-learn toolbox (Pedregosa *et al.*, 2011) and then assigned to the loss function during the model training process. A five-fold cross-validation evaluation was performed on each of the nine-class beta-turn classification tasks. The average experiment time ranged from 2 to 4 hours depending on the different amounts of data in each class.

**Table 5.** Nine-class beta-turn during balance learning, the performance was evaluated at turn-level on BT6376

| Beta-turn types | BetaTPred3 average MCC | DeepDIN Average MCC |
|---|---|---|
| Type I | 0.30 | 0.462 (±0.027) |
| Type I' | 0.45 | 0.645 (±0.057) |
| Type II | 0.35 | 0.569 (±0.050) |
| Type II' | 0.33 | 0.564 (±0.061) |
| Type IV | 0.20 | 0.292 (±0.016) |
| Type VIa1 | 0.33 | 0.395 (±0.023) |
| Type VIa2 | 0.25 | 0.262 (±0.121) |
| Type VIb | 0.35 | 0.465 (±0.023) |
| Type VIII | 0.17 | 0.234 (±0.015) |

## 4   Conclusion

In this work, a new deep neural network architecture DeepDIN was proposed for protein beta-turn prediction. Extensive experimental results show that DeepDIN obtained more accurate predictions than the best state-of-the-art methods and tools. Compared to previous machine-learning methods for protein beta-turn prediction, this work uses a more sophisticated, yet efficient, deep-learning architecture. There are several innovations in this work:

First, the proposed DeepDIN takes input of the entire protein sequence as the input feature, whereas all previous predictors relied on a fix-sized sliding window. DeepDIN is more effective and efficient in extracting long-range residue interactions. It utilizes densely connected inception blocks to effectively process local and non-local interactions of residues.

Second, we have implemented a tool called MUFold-BetaTurn which utilizes the proposed DeepDIN architecture for beta-turn prediction. This tool is ready for the research community to use freely.

Third, this work quantitatively discovered how different input features affect the beta-turn prediction. For example, some good features such as shape string and HHBlits profile can improve beta-turn classification effectively. In future work, the small beta-turn classes still have room for further improvement. Also, those features can be useful for other turn prediction problems, such as gamma-turn prediction (Pham *et al.*, 2005).

Last but not least, here we demonstrate a good example of the small imbalanced data set classification problem using balanced learning and transfer learning. The beta-turn data set is very imbalanced, where the ratio of positive samples over negative samples is about 1:9. Balanced learning and transfer learning were applied to overcome the problem. It is worth mentioning that transfer learning was originally applied to image recognition tasks, and here we applied a similar method to training models for small beta-turn classification tasks. The pre-trained network is effective in learning some more general beta-turn features; then the transfer learning technique can transfer the base network to some more specific models that can classify nine-class beta-turns. We have also demonstrated some techniques for tuning deep neural networks on small data classification problems, which may be useful in other areas of biological sequence analyses with imbalanced data sets such as genomic analysis, post translational modification prediction, etc.

## Acknowledgements

This work was partially supported by National Institutes of Health award R01-GM100701.

*Conflict of Interest:* none declared.